# From Data to Visualisations and Back: Selecting Visualisations Based on Data and System Design Considerations

Belgin Mutlu[1], Vedran Sabol[1], Heimo Gursch[2], Roman Kern[2]

Knowledge Visualization Group, Know-Center GmbH[1]
Knowledge Discovery Group, Know-Center GmbH[2]

**Abstract**

Graphical interfaces and interactive visualisations are typical mediators between human users and data analytics systems. HCI researchers and developers have to be able to understand both human needs and back-end data analytics. Participants of our tutorial will learn how visualisation and interface design can be combined with data analytics to provide better visualisations. In the first of three parts, the participants will learn about visualisations and how to appropriately select them. In the second part, restrictions and opportunities associated with different data analytics systems will be discussed. In the final part, the participants will have the opportunity to develop visualisations and interface designs under given scenarios of data and system settings.

## 1 Tutorial Content and Organisation

Creating high-quality user interfaces and visualisations requires understanding of interface design and visualisation guidelines. Since back-end data analytics and data storage impose restrictions on user interfaces and visualisations design, HCI researchers and developers have to be familiar with user interface design, visualisation selection and the features of the information processing system to create the best possible interface for all users. This tutorial covers context-sensitive choice and configuration of visualisations, as well as data analysis systems design. To that end, the tutorial is organised in three main parts. The first part the focuses on the state-of-the-art visualisation and interface design. We present theoretical background on perception, visual encoding rules, interaction design, and rules and guidelines for describing the semantics of visualizations. Based on these theoretical foundations, best practices and techniques are described to select visualizations. Using these guidelines, the participants can select new or evaluate existing designs based on data, system requirements, use case, target user groups, etc. To put the knowledge at work, we provide examples of cutting-edge interfaces and visualisations currently researched at the Know-Center, such as a personalised visualisation suggestion. Participants will be encouraged to assess the presented prototypes based on the above-mentioned guidelines and their personal experiences.

The second part of the tutorial focuses on data and the algorithms required to enhance visualisations. The participants will learn about various back-ends and their features with regard to interfaces and visualisations. Knowledge discovery algorithms will be discussed in connection with data storages, back-end, and visualisation requirements. The presented algorithms offer additional insights into data and extract relationships that are helpful in visualisations as well as provide user support in visual analytics. At the end of the second part, the interface designers should have an overview of the drawbacks and opportunities that various data storages provide and how to utilize these techniques in order to create more elaborate visualizations. Additionally, the participants will gain insights on what knowledge discovery algorithms can provide, especially if the raw data is not ideal for visualisations.

In the third part, small groups of participants will have to work on a given interface design and visualisation problem. They will receive a description of a use case and the data and system involved. Each group will have to develop an interface and visualisation design suited for the given problem. Poster material and pre-printed visual elements will be provided to the teams. The teams will have to design a poster illustrating their solutions. The teams will present their solution and collect feedback from the other teams and the tutors. The goal of this session is to master a real-life design scenario with the discussed information, guidelines and hand-outs from parts one and two. With the help of the tutors, the participants will be shown how such guidelines can be used in a day-to-day business of interface design.

## 2     Didactic Concept

Participants receive handouts describing the rules, guidelines and techniques presented during the tutorial. The tutorial will be divided into three sessions aligned with the three main parts presented above. Participants will be encouraged to ask questions and engage in discussions in the first two parts. In the third part, the tutors' attendants will form small groups with a mixed expertise in the fields of visualisations, interface design, and data analytics. In the mixed groups, the participants will share their experiences and knowledge to learn not only from the tutors but also from each other. This way, the tutorial will offer benefits to participants with various types of expertise.

In discussions, the presented material will be analysed based on the previous experiences of the tutorial participants. The goal of these discussions is to recap the presented material and put the knowledge to work as it is presented. The dissemination of knowledge will be organised in an interactive manner.

# 3  Target Audience and Number of Participants

This tutorial is targeted at researchers and developers of user interfaces and visualisations with different kinds of expertise. During the tutorial, groups will be formed and the participants will have to solve problems set by the tutorial's organisers. The organisers will mix the groups based on previous experience in the interface design and the visualisation domain. Each group will be supervised by a tutor. For optimal performance, the number of participants and groups will be limited to 25 participants and 5 groups.

The tutorial will be held in English. All presented slides and handouts will also be provided in English. The tutors speak English and German and will be able to answer questions any of these two languages. The individual teams may choose either English or German as a working language.

**Acknowledgments**

The Know-Center is funded within the Austrian COMET Program—Competence Centers for Excellent Technologies—under the auspices of the Austrian Federal Ministry of Transport, Innovation and Technology, the Austrian Federal Ministry of Economy, Family and Youth and by the State of Styria. COMET is managed by the Austrian Research Promotion Agency FFG.

**Authors**

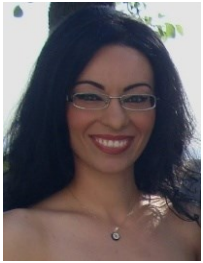

### Belgin Mutlu

Belgin Mutlu is currently pursuing her PhD's degree in Informatics at Know-Center GmbH and Graz University of Technology. She received her Master's and Bachelor's degree in Telematics both from Graz University of Technology. Her research interests include visual data analysis, recommender systems and semantic web.

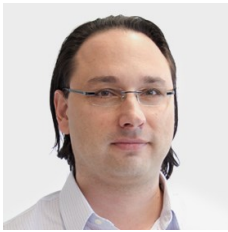

### Vedran Sabol

Vedran Sabol leads the Knowledge Visualization Area at the Know-Center. He is also employed at the Graz University of Technology as lecturer and senior researcher. His focus is the fields of visual analytics, information and knowledge visualization, and human-computer interaction. He has published over 60 peer-reviewed scientific publications.

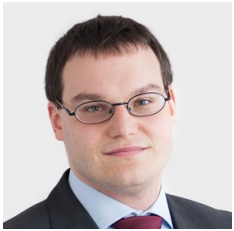

### Heimo Gursch

Heimo Gursch is working at the Knowledge Discovery division at the Know-Center. There he conducts research in the fields of information retrieval, sensor networks and data analytics. He holds a Master's degree in electrical engineering for the Technical University of Munich.

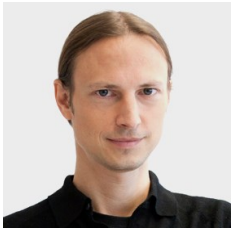

### Roman Kern

Roman Kern is the area manager of the Knowledge Discovery division at the Know-Center, where he works on machine learning, information retrieval and language processing. At the Graz University of Technology, he gives lectures and serves as supervisor for Bachelor, Master, and PhD students. He published over 50 peer-reviewed publications.